# Nitrogen-Vacancy Centers in Diamond for Current Imaging at the Redistributive Layer Level of Integrated Circuits

A. Nowodzinski[a, 1], M. Chipaux[b, *, 2], L. Toraille[b, **], V. Jacques[c], J-F. Roch[c], T. Debuisschert[b]

[a] *CEA, LETI, MINATEC Campus, F-38054 Grenoble France*
[b] *Thales Research & Technology, 1 av. Augustin Fresnel, 91767 Palaiseau Cedex, France*
[c] *Laboratoire Aimé Cotton, CNRS, Univ. Paris-Sud, ENS Cachan, Université Paris-Saclay, 91405 Orsay Cedex, France*

**Abstract**

We present a novel technique based on an ensemble of Nitrogen-Vacancy (NV) centers in diamond to perform Magnetic Current Imaging (MCI) on an Integrated Circuit (IC). NV centers in diamond allow measuring the three components of the magnetic field generated by a mA range current in an IC structure over a field of $50 \times 200$ µm$^2$ with sub-micron resolution. Vector measurements allow using a more robust algorithm than those used for MCI using Giant Magneto Resistance (GMR) or Superconducting Quantum Interference Device (SQUID) sensors and it is opening new current reconstruction prospects. Calculated MCI from these measurements shows a very good agreement with theoretical current path. Acquisition time is around 10 sec, which is much faster than scanning measurements using SQUID or GMR. The experimental set-up relies on a standard optical microscope, and the measurements can be performed at room temperature and atmospheric pressure. These early experiments, not optimized for IC, show that NV centers in diamond could become a real alternative for MCI in IC.

**Corresponding author:**
antoine.nowodzinski@cea.fr
Tel: +33 043878 1142

* current position: University of Groningen / University Medical Center Groningen, Department of Biomedical Engineering, Antonius Deusinglaan 1, 9713 AV Groningen, The Netherlands

** current position: Ecole Normale Supérieure de Lyon, 15 parvis René Descartes - BP 7000 69342 Lyon Cedex 07 – France

1 and 2: These two authors contributed equally to this work.

# Nitrogen-Vacancy Centers in Diamond for Current Imaging at the Redistributive Layer Level of Integrated Circuits

A. Nowodzinski, M. Chipaux, L. Toraille, V. Jacques, J-F. Roch, T. Debuisschert

## 1. Introduction

Magnetic Current Imaging (MCI) has demonstrated its ability to perform efficient failure localization in electron devices whether it is a "short" [1] or even "open" failure [2]. Commercial MCI for electron devices uses either Superconducting QUantum Interference Devices (SQUID) or Giant Magneto Resistance (GMR) sensor. Working at cryogenic temperature in a vacuum chamber, SQUID reaches sensitivity as low as 40 pT/Hz$^{1/2}$ [6] with microscale spatial resolution. On the other hand, GMR sensors allow measurement at room temperature and atmospheric pressure with a sensitivity close to 10 nT/Hz$^{1/2}$ [6] and sub-micron resolution. Magnetic field images obtained with those two techniques can be converted to current cartography thanks to the Biot-Savart law inversion as described by Roth [3].

Recently, the Nitrogen-Vacancy (NV) center in diamond has been studied for its remarkable properties. In particular, new types of magnetometer devices have been developed exploiting the spin properties of this color center. A first example is scanning field microscopy where a single NV center placed on an Atomic Force Microscope (AFM) tip, allows magnetic mapping at the nanoscale, with sensitivity down to 10 nT/Hz$^{1/2}$ [4]. A second example is wide-field magnetic imaging with an ensemble of NV centers located at the surface of a diamond plate. It directly provides a diffraction limited image, with a resolution of 500 nm, of the three spatial components of the magnetic field, with no requirement of any scanning procedure [5]. The measurement is shot-noise limited, which means that the minimum detectable magnetic field is inversely proportional to the square-root of the number of photons integrated over the time but also over the integration volume. Therefore, for a given NV layer thickness, the sensitivity has to be normalized over the integration surface and the integration time. A sensitivity of 2 µT.µm/Hz$^{1/2}$ has been reported [5].

The goal of this paper is to investigate the application of such wide-field magnetic imager for MCI. First, we describe the principle of magnetic imaging based on NV centers. Then, the measurement of the magnetic field produced by an IC is presented, as well as the MCI reconstruction.

## 2. Measurement of the magnetic field thanks to the NV color center in diamond

### 2.1. Principle

The NV center is a crystalline defect of diamond constituted of a nitrogen atom (N) substituted to a carbon atom and a vacancy (V) located on an adjacent crystalline site of the lattice (Figure 1.a).

The NV center is a perfectly photostable photoluminescent object with an internal degree of freedom related to its total spin angular momentum ($S = 1$). It can be optically polarized in its $m_s = 0$ state. Its spin state can be optically read out by its photoluminescence level that is lower for the $m_s = \pm 1$ than for the $m_s = 0$ spin state. Consequently, resonances between those spin states, induced by a microwave signal, can be detected by pure optical means.

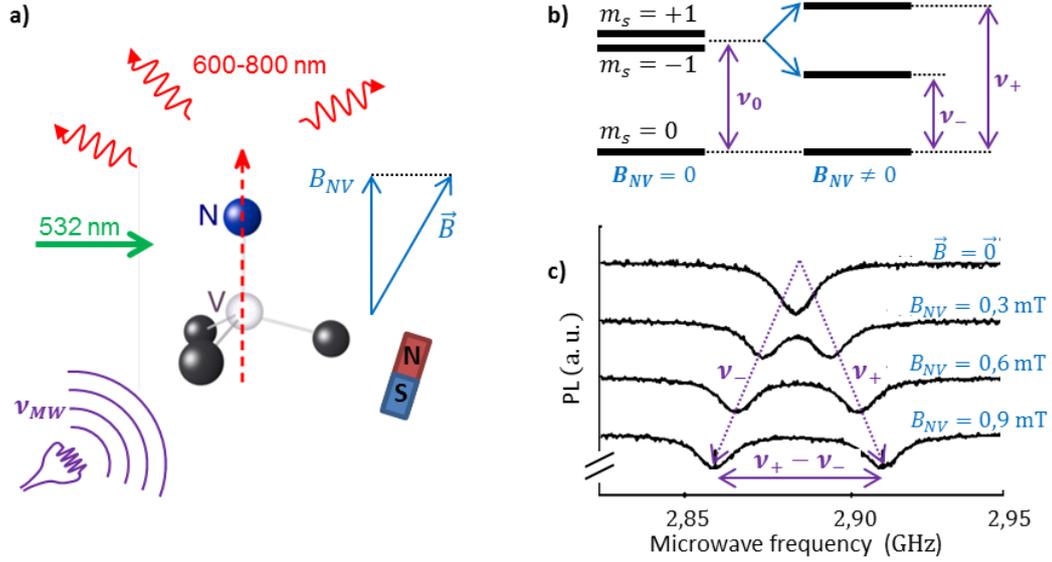

**Figure 1: NV center in diamond**
**a)** The NV center of diamond is constituted by a nitrogen atom (N), substituted to a carbon atom, and a vacancy in an adjacent site. This quantum object absorbs light in the green (at 532 nm in our case) and emits a perfectly stable photoluminescence in the red domain (between 600 to 800 nm). **b)** Energetic diagram associated to its internal electronic spin. The degeneracy between the zero spin state ($m_s = 0$) and the non-zero spin states ($m_s = \pm 1$) is lifted by the spin-spin interaction. In the presence of an external magnetic field (in blue), the degeneracy between the states $m_s = -1$ and the state $m_s = +1$ is lifted. The frequency difference, given by eq. (1) is proportional to the projection of the magnetic field on the NV axis. **c)** Electron Spin Resonances (ESR) spectrum. Resonances between the state $m_s = 0$ and the state $m_s = \pm 1$ induced by a microwave field (in purple) can be detected optically by a decrease of the photoluminescence.

Figure 1.c represents the photoluminescence of a NV center submitted to an increasing magnetic field. For a zero field, the states $m_s = -1$ and $m_s = +1$ are degenerated, and the spectrum presents only one resonance at $\nu_0 = 2,87$ GHz corresponding to the resonance between the states $m_s = 0$ and $m_s = \pm 1$. With a non-zero magnetic field, the degeneracy between the states of non-zero spin is lifted.

The two resonances, at $\nu_-$ and $\nu_+$ correspond to the transitions between the $m_s = 0$ state and the $m_s = -1$ and the $m_s = +1$ state respectively. The frequency difference is given by:

$$\nu^+ - \nu^- = 2\frac{g\mu_B}{h}B_{NV} \quad (1)$$

$B_{NV}$ is the projection of the magnetic field along the axis of the NV center, $g$ is the Landé factor of the NV center, $\mu_B$ is the Bohr magneton and $h$ is the Planck constant. The projection of the magnetic field along the NV axis can thus be deduced by the measurement of this frequency difference.

## 2.2. Wide-field magnetic imaging

The magnetic imager is described in detail in ref. 5. It relies on an ultrapure diamond plate holding an ensemble of NV centers located 10 nm below the surface. They are produced by ion implantation. As depicted in Figure 2.a, this layer is pumped by a green laser at 532 nm and submitted to a microwave field. An image of its luminescence signal is obtained with a standard optical microscope. (Figure 2.b)

Here, the field of view (around $50 \times 200$ μm$^2$) is determined by the size of the laser spot. In our case, the use of a microscope objective with a magnification of 10 and a numerical aperture of 0.1 is well adapted to the object we are investigating.

A microwave excitation is applied to the sample. Its frequency is swept in the range around 2.87 GHz and an image of the luminescence over the region of interest is taken at each step. Consequently, a whole ESR spectrum can be obtained for each single pixel of the image (Figure 2.c).

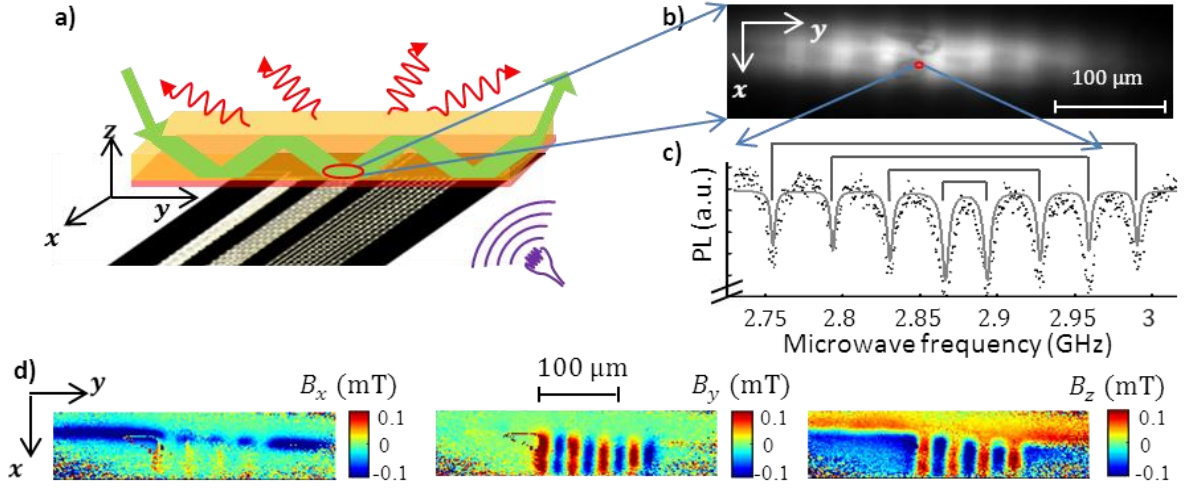

**Figure 2: Wide-field magnetic imaging with an ensemble of NV centers**
**a)** An ultrapure 4 mm × 4 mm × 250 μm diamond plate (yellow) holds an ensemble of NV centers (red) located close to its main face. This active layer is placed in contact to the sample (black). Then, a pumping beam (green) is propagated inside the diamond and illuminates an area of this active layer. A microwave field is applied through an antenna (purple). The photoluminescence signal is collected in direction $z$ by a standard optical microscope and imaged on a digital camera. **b)** Image of the photoluminescence of the NV centers layer. **c)** ESR spectrum for one given pixel. **d)** Spatial components of the magnetic field calculated from the images of the photoluminescence.

Each spectrum features four pairs of resonances. They correspond to the projections of the magnetic field along the four possible NV center axis orientations inside the diamond lattice (Eq. (1)). Exploiting those four projections allows retrieving the three vector components of the magnetic field.

### 3. Magnetic Current Imaging

#### 3.1. Principle

In ref. 3, Roth describes how to reverse the Biot-Savart law and thus calculate the current density when the magnetic field is measured. It is shown that $B_x$, $B_y$ and $B_z$, the three components of the magnetic fields, are the results of a convolution between filters and the two components of the current in the plane: $J_x$ and $J_y$ (the vertical component is assumed to be zero).

GMR and SQUID sensors are only sensitive to $B_z$ which is itself dependent on two components $J_x$ and $J_y$. Thanks to the addition of the hypothesis of continuity of current, the current image may be reconstructed since we have a system of two equations with two unknown variables ($J_x$ and $J_y$).

NV color centers also provide $B_x$ and $B_y$, the two other components of the magnetic field, which thus builds an over-determined system of four equations and two unknown variables. This adds constraints on the current image and therefore should reduce the noise.

In ref. 2, the Biot-Savart law is written as an integral of a vector product where $\mu_0$ is the vacuum permeability and $\vec{j}$ is the current density that generates the magnetic field $\vec{B}$ at a position defined by the vector $\vec{r}$.

$$\vec{B}(\vec{r}) = \frac{\mu_0}{4\pi} \int \frac{\vec{j}(\vec{r'}) \wedge (\vec{r}-\vec{r'})}{|\vec{r}-\vec{r'}|} d^3\vec{r'} \qquad (2)$$

It can be viewed as the convolution of the current density with a filter function. Since $z$ is fixed to $z_{mes}$, it is a two-dimensional problem in the $x$ and $y$ directions of space. It takes a simpler expression in the ($k_x$, $k_y$) Fourier space:

$$\vec{B}^{TF}(k_x, k_y, z_{mes}) = \frac{\mu_0}{4\pi} \vec{j}^{TF} \wedge \vec{F} \qquad (3)$$

where $\vec{B}^{TF}$ and $\vec{j}^{TF}$ are the Fourier transforms of the magnetic field and current density respectively. $\vec{F}$ is the Fourier transform of the

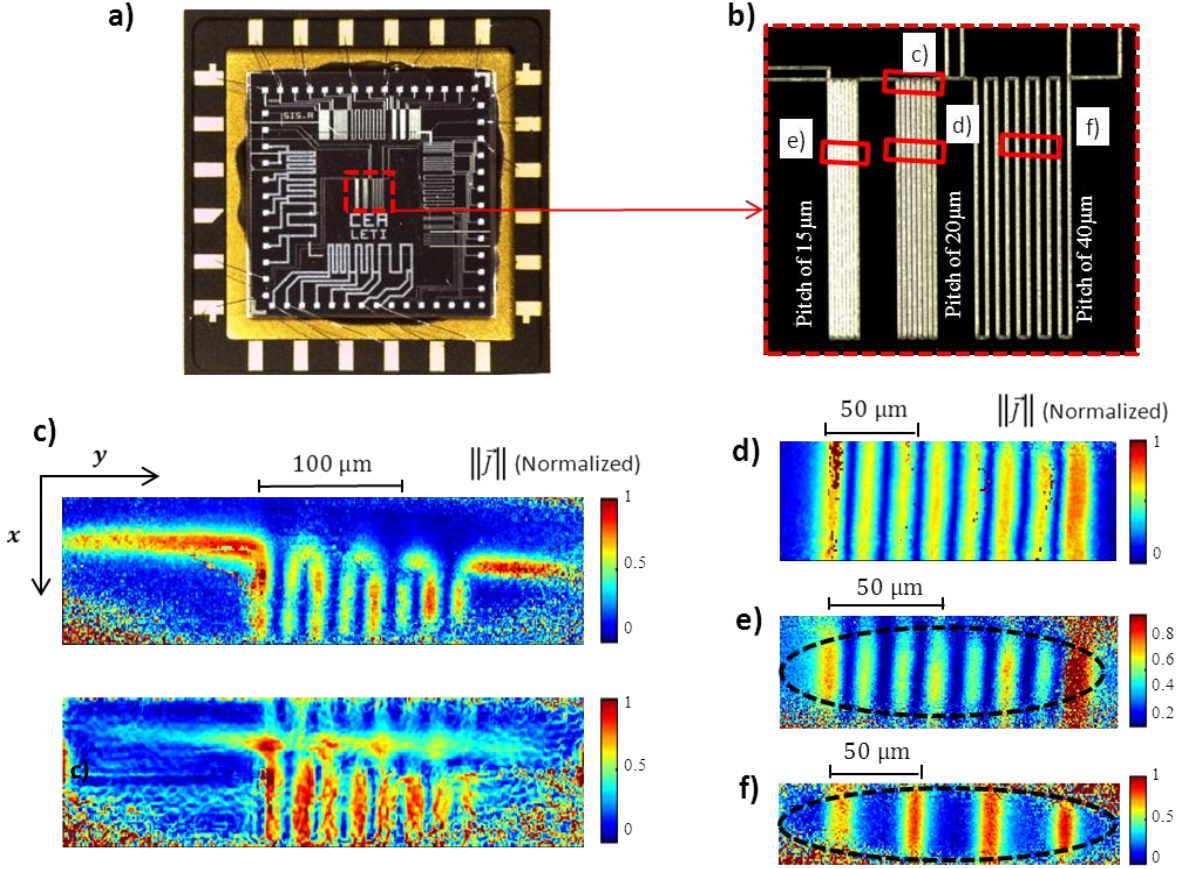

**Figure 3**: **MCI of a test sample. Comparison of MCI using $B_z$ and MCI using $B_x$, $B_y$ and $B_z$**
**a)** View of the integrated circuit used for the test, the dashed line encloses the area where the measurements were made. **b)** Details of the area used for the measurements. Three serpentines are used for MCI testing, for all of them the line is 10 μm wide, but the pitches are 15 μm, 20 μm and 40 μm respectively. **c)** MCI of a current of 6 mA using the three components of the magnetic field (upper figure) or using only the component in the z direction of the space (lower figure), in the latter case the serpentine is hardly recognizable **d)** and **e)** MCI of a current of 10 mA. **f)** MCI of a current of 8 mA. The dashed lines delimit the area of the laser spot, outside of which the NV centers are not efficiently pumped.

vector $\vec{f}$ whose components are given by:

$$f_l(x, y, z_{mes}) = \frac{l}{\sqrt{x^2+y^2+z_{mes}^2}} \quad (4)$$

for $l = x, y, z_{mes}$. In addition to the expressions of the components of the magnetic field, one can use the continuity of current expressed in the frequency domain by:

$$i.k_x . J_x^{TF} + i.k_y . J_y^{TF} = 0 \quad (5)$$

where we have considered an in-plane current ($J_z^{TF} = 0$). The over-determined system of 4 equations and two unknowns ($J_x$ and $J_y$) can be expressed in matrix form by:

$$B = M * J \quad (6)$$

with

$$B = \begin{vmatrix} B_x^{TF} \\ B_y^{TF} \\ B_z^{TF} \\ 0 \end{vmatrix}; \quad M = \begin{vmatrix} 0 & F_{z_{mes}} \\ -F_{z_{mes}} & 0 \\ F_y & -F_x \\ i.k_x & i.k_y \end{vmatrix}; \quad J = \begin{vmatrix} J_x^{TF} \\ J_y^{TF} \end{vmatrix} \quad (7)$$

In this paper, standard matlab@ QR solver is used to find the matrix of current density $J^{TF}$. Once the two components of the current density are found in the space frequency, an inverse FFT is applied to obtain the current in the real space.

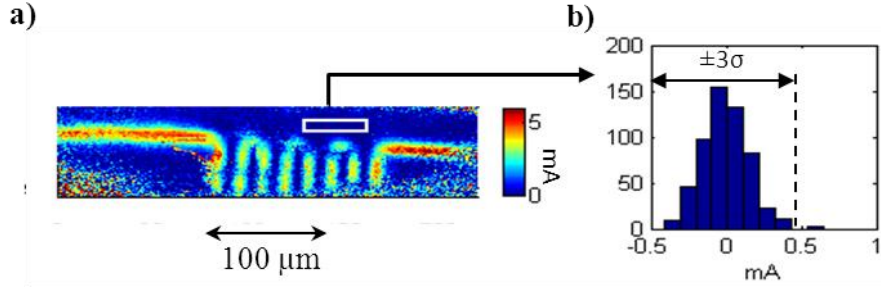

**Figure 4**: **Evaluation of the set-up sensitivity**
**a)** Fig 3c : Current reconstruction after offset subtraction and normalization to 6 mA. **b)** Histogram of the area surrounded by a white line in the fig.4a. The measured standard deviation is 0.15 mA, it was thus considered that the minimum detectable current is 0.5 mA (≈3σ). It is important to note that this threshold is dependent on the distance between the sensor and the sample, and on the geometry of the conductive line.

Here, the evaluation of MCI, calculated from NV color centers, is only a 2D structure. It is worth noting that the system of equations (7) can be transformed into a system of 4 equations with 3 unknowns ($J_x^{TF}$, $J_y^{TF}$ and $J_z^{TF}$) in order to calculate the 2D image of $J_z^{TF}$, the out-of-plane component of a current flowing in a 3D structure.

It is clear that the vector magnetic field measurement authorized by the use of NV color centers opens new current reconstruction prospects.

*3.2. Measurements*

The sample used for MCI is depicted on Figures 3a and 3b. It consists of metal lines deposited over resin and forming serpentines. One of these serpentines is far enough from the pads to accommodate the size of the diamond (4x4 mm²).

Figures 3c) to 3f) show the good agreement between assumed current path and calculated MCI. On Figures 3e) and 3f), the dashed lines delimit the area of the laser spot, outside of which the NV centers are not efficiently pumped.

For each MCI, the magnetic field acquisition time is 10 s and the calculation time for the reconstruction of the magnetic field is 1 minute. This very short measurement time should be compared with the case of sensor scanning instruments whose acquisition time may exceed several tens of minutes.

Figure 3.c shows the interest of calculating the MCI from the vector magnetic field rather than just the vertical component of the magnetic field as it is done with classical MCI. To calculate the MCI with only the vertical component, we use the same matrix B and M but remove the first two lines of M and B in order to build a system of 2 equations and 2 unknowns ($J_x^{TF}$, $J_y^{TF}$). Moreover, a Blackman window is applied before calculating the inverse FFT of the current density components. This operation is not necessary when the MCI is calculated from the three components of the magnetic field but it helps to suppress the high frequency noise generated by the current step at the edge of the frequency domains of $J_x^{TF}$ and $J_y^{TF}$.

The quality difference between the MCI calculated from the three components of the magnetic field and the MCI calculated from only the vertical magnetic field can be explained because the over-determined system defined in (7) provides redundant information to calculate the current density. Thus, the MCI obtained from the three components of the magnetic field is much more tolerant to measurement noise than the MCI obtained with only the vertical component of the magnetic field. Besides, the sensor is closer to the current, which is one of the reasons why, despite large measurement noise (compared with GMR or SQUID sensors), MCI remains feasible at the milliampere range.

Figure 4 is an attempt to determine the minimum detectable current for the sample and for the measurement configuration used in this paper. It has been found that the minimum detectable current is 0.5 mA , which is very promising for a first non-optimized setup and with an acquisition time of 10 s.

## 4. Conclusion

The MCI described in this paper with a setup not optimized for IC shows remarkable results. The instrument providing the vector magnetic field for the MCI calculation is very simple and usable at room temperature and atmospheric pressure. It enables measuring the magnetic field vector over a field of view of $50 \times 200\ \mu m^2$ in less than 10 s. The field of view is limited by the laser spot size. It can be easily expanded enlarging the laser spot. The surface of the diamond plate can be reduced to perform measurement on smaller IC. Then the spatial resolution can be decreased down to 500 nm, at the limit given by optical diffraction.

Vector measurements allow using a more robust algorithm than those used for MCI using GMR or SQUID sensors and may suggest that new reconstruction algorithm could be developed to characterize currents in 3D structures.

With the setup and sample used in this paper the current resolution is estimated to be 0.5 mA. Several ways to improve it can be investigated. For example, differential measurements can eliminate the background noise [5], which would directly result in an improvement of the sensitivity and allow a total acquisition and treatment time smaller than one second. In addition, several ways to improve the performances of the wide-field magnetic imager itself can be foreseen, using either optimized diamond crystals [7], [8] or pulsed measurement techniques [9], [10].

## Acknowledgements

The magnetic field measurements were performed in Thales Reasearch & Technology (Palaiseau, Fr.). The MCI reconstruction was performed in CEA-Leti (Grenoble, Fr.). The research leading to these results has received funding from the European Union Seventh Framework Programme (FP7/2007-2013) under the project DIADEMS (Grant agreement No. 611143) and from the Agence Nationale de la Recherche (ANR) under the project ADVICE (Grant ANR-2011-BS04-021).

## References

[1] F. Infante et al., A new Methodology for Short Circuit Localization on Integrated Circuits using Magnetic Microscopy Technique Coupled with Simulations IEEE Proceedings of 16th IPFA - 2009, China

[2] Jan Gaudestad et al., Space Domain Reflectometry for Open Failure Localization), 2012 19th IEEE International Symposium on the Physical and Failure Analysis of Integrated Circuits (IPFA)

[3] Bradley J. Roth et al., Using a magnetometer to image a two dimensional current distribution *J. Appl Phys.* **65** (1), 1 January 1989

[4] L Rondin et al., Magnetometry with nitrogen-vacancy defects in diamond, *Rep. Prog. Phys.* **77** (2014)

[5] M. Chipaux et al., Magnetic imaging with an ensemble of Nitrogen Vacancy centers in diamond, Eur. Phys. J. D (2015) **69**: 166.

[6] http://www.neocera.com/magma/Products.html, commercial brochure

[7] M. Lesik, T. Plays, A. Tallaire, J. Achard, O. Brinza, L. William, M. Chipaux, L. Toraille, T. Debuisschert, A. Gicquel, J.F. Roch, V. Jacques, Preferential orientation of NV defects in CVD diamond films grown on (113)-oriented substrates, Diamond and Related Materials, Volume 56, June 2015, Pages 47-53, ISSN 0925-9635.

[8] M. Lesik, J.P. Tetienne, A. Tallaire, J. Achard, V. Mille, A. Gicquel, J.F. Roch, V. Jacques, Perfect preferential orientation of nitrogen-vacancy defects in a synthetic diamond sample. Appl. Phys. Lett., 104 (2014) 113107.

[9] J. M. Taylor, P. Cappellaro, L. Childress, L. Jiang, D. Budker, P. R. Hemmer, A. Yacoby, R. Walsworth et M. D. Lukin, «High-sensitivity diamond magnetometer with nanoscale resolution», Nat. Phys, vol. 4, n° 110, pp. 810-816,2008.

[10] A. Dréau, M. Lesik, L. Rondin, P. Spinicelli, O. Arcizet, J.-F. Roch et V. Jacques, «Avoiding power broadening in optically detected magnetic resonance of single NV defects for enhanced dc magnetic field sensitivity», Phys. Rev. B, vol. 84, n° 119, p.195204, 2011.